\documentclass[twocolumn]{aastex631}
\usepackage{amsmath,amsfonts,amssymb}
\usepackage{bm}
\usepackage{graphicx}

\received{Submission Date}
\submitjournal{\psj}

\begin{document}
\graphicspath{ {/rapid/cproj/haumea/figures/} }

\title{Spatially variable crater morphology on the dwarf planet Haumea}

\author{George D. McDonald\altaffiliation{1}}
\affiliation{Department of Earth and Planetary Sciences, Rutgers, The State University of New Jersey, Piscataway, NJ, USA}
\affiliation{Department of Electrical \& Computer Engineering, Portland State University, Portland OR, USA}

\author{Lujendra Ojha}
\affiliation{Department of Earth and Planetary Sciences, Rutgers, The State University of New Jersey, Piscataway, NJ, USA}

\correspondingauthor{George D. McDonald\altaffiliation{1}}
\email{georgem@pdx.edu}

\begin{abstract}
  Haumea, thought to be the Kuiper Belt's 3rd most massive object, has a fast 3.92 hr rotational period, resulting in its shape as a triaxial ellipsoid. Here we calculate Haumea's surface gravity field and make the first detailed predictions of Haumea's surface morphology. We focus on crater characteristics, with craters likely the predominant surface feature, considering infrared spectroscopy has indicated Haumea's surface to be predominantly inert water ice. In calculating Haumea's surface gravity, we find that $g$ varies by almost two orders of magnitude, from a minimum of 0.0126 m/s$^2$ at the locations of the equatorial major axis, to 1.076 m/s$^2$ at the poles. We also find a non-monotonic decrease in $g$ with latitude. The simple to complex crater transition diameter varies from 36.2 km at Haumea's locations of minimum surface gravity to 6.1 km at the poles. Equatorial craters are expected to skew to larger volumes, be at least 2$\times$ times deeper, and have thicker ejecta when compared with craters at high latitudes. Considering implications for escape of crater ejecta, we calculate that Haumea's escape velocity varies by 62\% from equator to pole. Despite higher escape velocities at the poles, impacts there are expected to have a higher mass fraction of ejecta escape from Haumea's gravitational well. Haumea may be unique among planet-sized objects in the solar system in possessing dramatic variations in crater morphology across its surface, stemming solely from changes in the magnitude of its surface gravity.
\end{abstract}

\section{Introduction}
The dwarf planet Haumea is the third brightest \citep{brown2006} and at 4.006 $\pm$ 0.040 $\times$ 10$^{21}$ kg, the third most massive Kuiper Belt Object \citep{ragozzine2009,rambaux2017,dunham2019}. \footnote{The uncertainty on Makemake's mass allows for a small possibility that it is more massive, which would make Haumea 4th.} Early after its discovery, Haumea was determined to be an extraordinary object. Its $\sim$3.92 hr rotation period is the shortest among solar system objects larger than 100 km and thought to be the result of a giant impact collision, either catastrophic \citep{brown2006,brown2007} or a graze-and-merge collision \citep{leinhardt2010,proudfoot2019, proudfoot2022,noviello2022}. The rapid rotation rate imparted by Haumea's formation was thought to result in its shape being either a triaxial ellipsoid or an oblate spheroid where the equatorial and polar axes  differed by $>$ 30 \% \citep{rabinowitz2006}. Later, photometric and thermal flux measurements confirmed that Haumea was indeed a triaxial ellipsoid \citep{lockwood2014}. The most precise dimensions of Haumea come from the stellar occultation observations of \citealt{ortiz2017} with equatorial axes of $a$ = 1161 $\pm$ 30 km and $b$ = 852 $\pm$ 4 km, and a polar axis of $c$ = 513 $\pm$ 16 km.

\begin{figure*}
  \centering
  \epsscale{1.15}
  \plotone{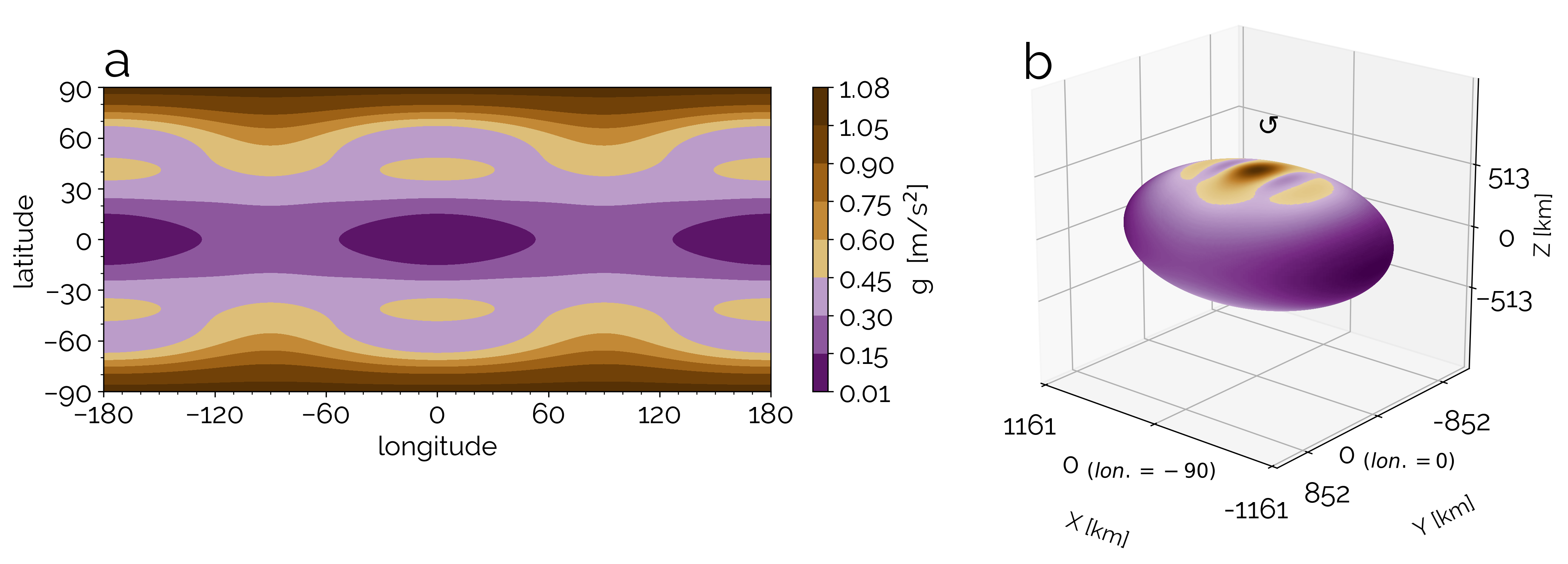}
  \caption{(a) The magnitude of the surface gravitational acceleration $g$ as a function of latitude and longitude, in an equirectangular projection. Purple corresponds to low gravity and brown corresponds to high. (b) The same as (a), but in a 3 dimensional perspective. The axes here show distances corresponding to the lenths of the axes, rather than longitude, with longitudes of -90$^{\circ}$ and 0$^{\circ}$ labeled for orientation. The circular arrow depicts Haumea's direction of rotation.}
  \label{fig1}
\end{figure*}

Spectroscopic and photometric observations have provided valuable information about Haumea's surface. Infared spectroscopy by \citealt{trujillo2007} indicated a surface composition of 66 $\sim$ 80\% crystalline water ice, while \citealt{pinilla-alonso2009} used infrared spectroscopy in concert with Hapke scattering models to favor a surface covered by $>$ 92\% water ice, in close to a 1:1 ratio of amorphous to crystalline ice. Haumea's largest satellite, Hi'iaka, shares this water ice composition \citep{barkume2006}. The presence of a heterogeneous surface in the form of a ``dark red spot'' was indicated by photometry, although the cause for this is unknown. While a distinct composition for this region is thought to be more likely, it may also be explained by variations in the water ice grain size \citep{lacerda2008}. However, to date, there exists no method to observationally constrain the surface morphology of Haumea, and no studies have made detailed predictions for what might be expected of Haumea's surface morphology.

In recent years, a major advancement in our knowledge of Kuiper Belt Object surfaces was made by the observations of the \textit{New Horizons} mission. These observations revealed the dwarf planet Pluto to be a complex world--possessing both recently active geologic processes \citep{moore2016}, and extensive atmospheric photochemistry \citep{gladstone2016}. Pluto's largest satellite, Charon, while posessing an older and largely cratered surface, also provided evidence for endogenic activity possibly related to an internal ocean and cryovolcanism \citep{moore2016}. \textit{New Horizons} also provided new, high resolution observations of impact craters for two icy bodies.

The availability of findings from \textit{New Horizons}, in addition to the existing constraints on Haumea, make it timely to theorize on possible surface morphologies for this dwarf planet. Haumea's surface is predominantly water ice, which barring substantial present-day internal heat, will be involatile in the Kuiper belt \citep{brown2011}. The lack of any substantial volatile component at the surface of Haumea precludes the mass movement of glacial flows akin to Pluto's Tombaugh Regio \citep{moore2016}, as well as any substantial vapor pressure supported atmosphere \citep{gladstone2016}. An upper limit of 3 -- 50 nbar on the pressure of any atmosphere, depending on composition, has also been provided by \citealt{ortiz2017}. The crystalline nature of the water ice suggests some sort of communication between the surface and interior, as amorphous water ice is more energetically favorable at Kuiper Belt conditions and in the absence of a substantial atmosphere or magnetosphere, radiation will convert crystalline ice to amorphous ice on timescales of $\sim$10$^7$ yr \citep{cooper2003,jewitt2004,trujillo2007}. \citealt{pinilla-alonso2009} favor outgassing or the exposure of fresh material from large impacts (rather than cryovolcanism, due to the similar composition of the much smaller Haumea group objects), and estimate the surface age to be $>$ 10$^8$ yr.

We note several processes observed on Charon and the Saturnian satellites that are likely to be operating on Haumea. This includes the tectonics that can result from the freezing of a subsurface ocean \citep{moore1983,moore1985}, including potential associated cryoflows \citep{beyer2019} . Haumea has been proposed to have possessed a subsurface ocean for $\sim$250 Myr \citep{noviello2022}. Mass wasting is another process likely present on Haumea, as it can be triggered by impact shaking even in the absence of volatiles \citep{singer2012,fassett2014}. We nevertheless expect, that as with Charon and the mid-sized Saturnian satellites, an older surface coupled with few volatiles will make impact cratering a dominant surface process expressed across the entirety of Haumea's surface, which we focus on in this work.

We first quantify Haumea's surface effective gravity and its spatial variations, which stem from Haumea's unique shape and rapid rotation. This variable surface gravity drives the trends in the subsequent phenomena that we examine. We look at predicting crater types (simple vs complex) and morphometries (dimensions), and how they vary across Haumea's surface. We then look at spatial variations in crater ejecta characteristics. Ejecta on icy bodies, particularly the Callisto and Ganymede, exhibit a wide variet of characteristics (including high albedos, plateaus and scarps), and are thus of interest. Finally, we look at how the fraction of ejecta that can escape from Haumea's gravitational well varies across the surface.

\begin{figure}
  \centering
  \epsscale{1.2}
  \plotone{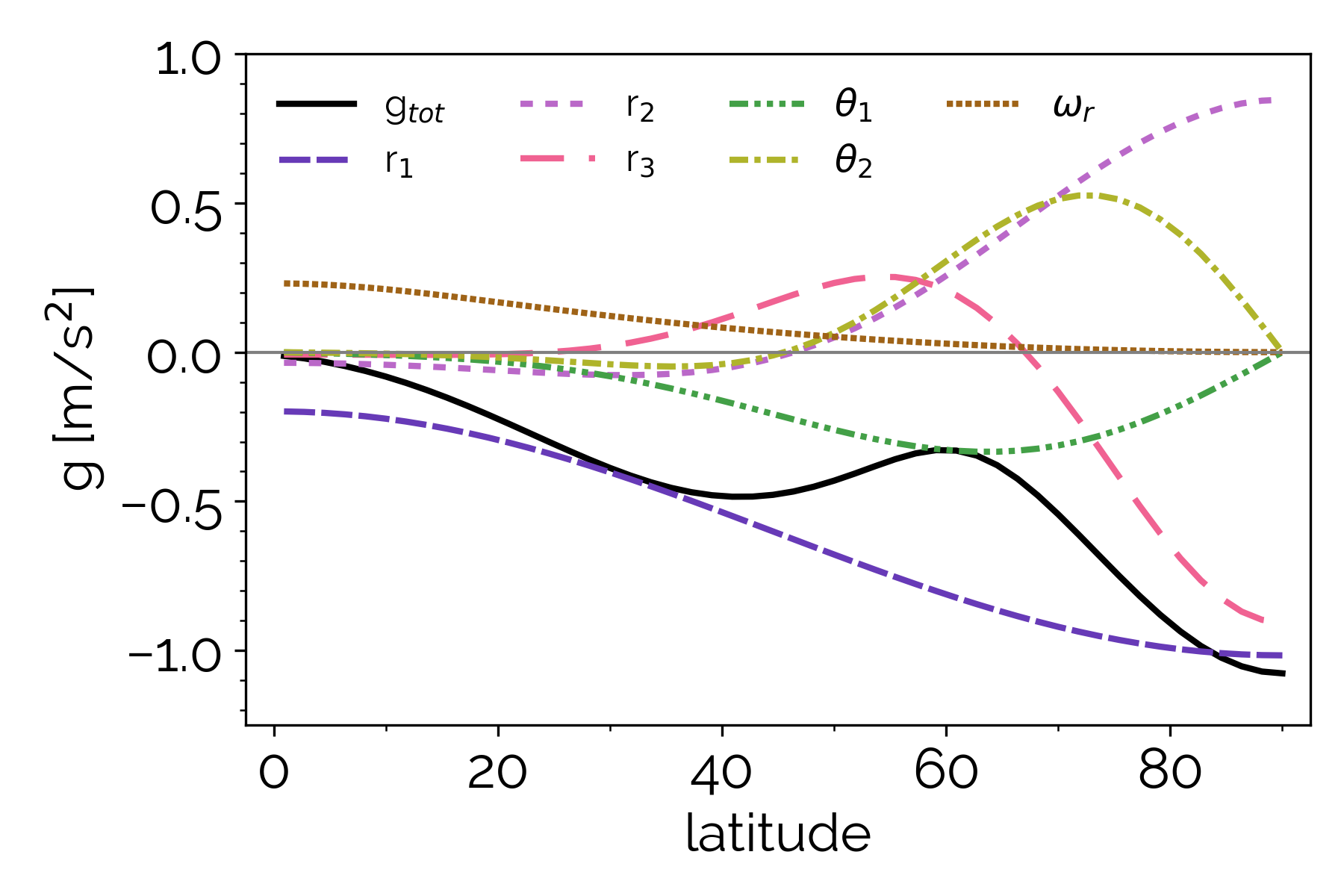}
  \caption{The magnitude and sign of the total surface gravitational acceleration $g$ as a function of latitude, at longitude = 0$^{\circ}$. The values of the 6 largest terms that contribute to the total magnitude of $g$ are also shown, with signs indicated to show their relative contributions during summation within each $g$ component.}
  \label{fig2}
\end{figure}

\section{Haumea's surface gravity}
\label{gravity}
\subsection{Coordinate system}
\label{gravity:coord}

Throughout calculations in the manuscript, we adopt spherical coordinates with radial distance $r$, polar angle $\theta$, and azimuthal angle $\lambda$. For plotting and geographic interpretation, $\theta$ and $\lambda$ are converted to latitude and longitude respectively. We define 0$^{\circ}$ azimuth and longitude to align with Haumea's equatorial semi-major axis, $a$. By this convention, 180$^{\circ}$ longitude also corresponds with the equatorial major axis, while the equatorial minor axis corresponds to longitudes of -90 and 90$^{\circ}$. Our adopted longitudes are positive eastward. With this convention, the direction of increasing longitude aligns with the direction of Haumea's rotation.  Spherical coordinates provide computational convenience, but we note that their use in specifying the surface of a triaxial ellipsoid results in some peculiarities. Specifically, both the latitudinal and longitudinal angles subtended by the same arc length will vary as a function of location on Haumea. In order to help orient the reader with respect to these effects, spatial plots are shown as both an equirectangular projection with latitude and longitude coordinates, as well as a three dimensional perspective with axes in units of km (see Figure \ref{fig1}).

\subsection{Methods: Gravity}
\label{gravity:mthd}

The effective gravity potential at Haumea's surface is the sum of the gravitational and centrifugal potentials. The effective gravity potential is expressed as a series of spherical harmonics.

\begin{align}
\begin{split}
  \label{eqn1}
  \Phi(r,\theta,\lambda) = - \frac{GM}{r} \biggl\{
                             1 & + \sum_{n=2}^{\infty} \sum_{m=0}^{n}
                             \left( \frac{R_o}{r} \right)^n P_n^m (\cos \theta) \\
                           \times [ & C_{nm} \cos m \lambda +  S_{nm} \sin m \lambda] \biggr\} \\
                           & - \frac{1}{2} \omega^2 r^2 \sin^2 \theta
\end{split}
\end{align}

where $G$ is the univeral gravitational constant, $M$ is Haumea's mass, $R_o$ is the mean radius, and $P_n^m$ are the associated Legendre polynomials. The summations are over the degree, $n$, and order, $m$, of the polynomials. $\omega$ is the angular velocity from Haumea's rotation. $C_{nm}$ and $S_{nm}$ are the spherical harmonic coefficients. For a triaxial ellipsoid, due to symmetries, $S_{nm}$ = 0 for all $n$ and $m$, while specifically due to north-south symmetry $C_{nm}$ = 0 for all odd $n$ and $m$. We evaluate the gravitational potential to the 4th order. We use the coefficients $C_{20}$ through $C_{44}$ as calculated by \citealt{sanchez2020}, who used Haumea's shape as determined by \citealt{ortiz2017}, and the methodology of \citealt{balmino1994} for calculating the coefficients. \citealt{balmino1994} present a method for calculating the spherical harmonic gravity coefficients for a triaxial ellipsoid, assuming a homogeneous composition (i.e. uniform density).

We then calculate the effective surface gravitational acceleration (hereafter surface gravity) as the negative of the gradient of the effective gravity potential.

\begin{align}
\begin{split}
    \label{eqn2}
    \vec{g} = & - \vec{\nabla} \Phi \\
            = & - \frac{\partial \Phi}{\partial r} \hat{r}
                - \frac{1}{r} \frac{\partial \Phi}{\partial \theta} \hat{\theta}
                - \frac{1}{r \sin \theta} \frac{\partial \Phi}{\partial \lambda} \hat{\lambda}
\end{split}
\end{align}

In evaluating the gravitational acceleration at the surface of Haumea, we need to calculate the distance to the origin at given coordinates ($\theta$,$\lambda$) on Haumea's surface. We do this by using the equation for a triaxial ellipsoid in spherical coordinates

\begin{align}
\begin{split}
    \label{eqn3}
    \frac{r^2 \cos^2\theta \sin^2\lambda}{a^2} +
    \frac{r^2 \sin^2\theta \sin^2\lambda}{b^2} + \frac{r^2 \cos^2\lambda}{c^2}
    = 1
\end{split}
\end{align}

and solving for $r$($\theta$,$\lambda$). The physical parameters that we adopt for Haumea, as well as the numerical constants used in all calculations in the manuscript are summarized in Table \ref{table1}.

\subsection{Results: Gravity}
\label{gravity:rslt}

Figure \ref{fig2} shows the sign and total magnitude of the surface gravity $g$ as a function of latitude at 0$^{\circ}$ longitude, showing an increase with latitude. The 6 terms with the largest magnitudes contributing to $g$ are also plotted. The terms are labeled according to the convention illustrated for the $\vec{g_r}$ terms below

\begin{align}
\begin{split}
  \label{eqn4}
  \vec{g_r} = - \frac{\partial \Phi}{\partial r} \hat{r}
              = ( g_{r,1} + g_{r,2} + g_{r,3} + \omega_r ) \hat{r}
\end{split}
\end{align}

where on the plot itself $g_{r,1}$ is labeled as $r_1$ for visibility. For the full expansion the individual terms, as well as the other surface gravity components, the reader is referred to the Appendix. Note that the $\vec{g_r}$ and $\vec{g_\theta}$ terms have different directions, and that furthermore the unit vector $\hat{\theta}$ varies as a function of location. The individual terms are shown to demonstrate which terms are contributing greatest to the total magnitude of the surface gravitational force $g$. The signs of individual terms are also shown, as these are summed together and various portions negate each other before the root mean square of the individual components is taken to calculate $g$. Lastly, we plot $g$ with a negative sign because despite its direction changing over Haumea's surface, it is always closer to pointing radially inwards vs. outwards.

\begin{deluxetable}{cccc}
\tabletypesize{\footnotesize}
\tablehead{
\colhead{Name} & \colhead{Value} & \colhead{Description} & \colhead{Reference}
}
\decimalcolnumbers
\startdata
\multicolumn{4}{l}{Haumea physical properties} \\
\hline
$\rho$ & 1885 kg/m$^3$                & Uniform density     & a        \\
$a$    & 1161 km                      & Equatorial semi-major axis & a \\
$b$    & 852 km                       & Equatorial semi-minor axis & a \\
$c$    & 513 km                       & Polar semi-axis   & a          \\
$R_o$  & 797.6 km                     & Mean radius       & b, c       \\
$\omega$ & 4.457 $\times$ 10$^{-4}$   & Angular velocity  & d           \\
$C_{20}$ & -0.114805                  & Spherical harmonic coeff. & e \\
$C_{22}$ & 0.230731 $\times$ 10$^{-1}$ & & \\
$C_{40}$ & 0.305251 $\times$ 10$^{-1}$ & & \\
$C_{42}$ & -0.189209 $\times$ 10$^{-2}$ & & \\
$C_{44}$ & 0.950665 $\times$ 10$^{-4}$ & & \\
\hline
\multicolumn{4}{l}{Imapact related parameters} \\
\hline
$\delta$ & 930 kg/m$^3$               & Impactor density (ice)    & f \\
$Y$      & 1.5 $\times$ 10$^7$ Pa     & Target strength  (ice)    & f \\
$K_1$    & 0.06                       & Volume scaling const. (hard rock) & f \\
$K_2$    & 1                          & Strength scaling constant (ice) & f \\
$\mu$    & 0.55                       & Scaling exponent (ice) & f \\
$\nu$    & 0.33                       & Scaling exponent (ice) & f \\
$K_r$    & 1.1                        & Simple crater diameter const. (ice) & f \\
$\alpha_E$ & 0.6117                   & Ejecta scaling exponent (ice) & g\tablenotemark{*} \\
$K_{vg}$ & 3.3                        & Ejecta velocity exp. (dense sand) & h \\
\enddata
\caption{Adopted values for physical parameters and numerical constants throughout calculations in the manuscript.}
\tablerefs{\textsuperscript{a}\citealt{ortiz2017}      \textsuperscript{b}\citealt{dunham2019}
           \textsuperscript{c}\citealt{kondratyev2020} \textsuperscript{d}\citealt{lellouch2010}
           \textsuperscript{e}\citealt{sanchez2020}    \textsuperscript{f}\citealt{holsapple2022}
           \textsuperscript{g}\citealt{senft2008}       \textsuperscript{h}\citealt{holsapple2012}
}
\tablenotetext{*}{Derived from fitting to the data in this reference}
\label{table1}
\end{deluxetable}

The overall trend is for $g$ increasing with latitude. The reason for this is analogous to that on Earth---flattening from Haumea's rotation results in a shorter polar axis vs the equatorial axes. This is coupled with the centrifugal acceleration increasingly opposing the gravitational acceleration at lower latitudes. At the equator, $g_{r1}$ and $\omega_r$ are comparable in magnitude, with values of -0.2 and +0.231 respectively. The consequence of this is an extremely low surface gravity at the equator of -0.0126 m/s$^2$ (the negative sign indicates direction). Several other features warrant discussion. The overall strength of the $g_{r,3}$ term, combined with a local maximum at 55$^{\circ}$ latitude and a change in sign at 67$^{\circ}$ latitude contribute to $g$ not increasing monotonically with latitude. Specifically, this results in a local minimum in $g$ at 42$^{\circ}$, and a local maximum at 60$^{\circ}$ latitude. Thus, there are degeneracies on Haumea in that multiple latitudes within the same hemisphere can have the same $g$ value. While the $g_{\theta}$ terms largely cancel each other out below 60$^{\circ}$ latitude, from 60 -- 85$^{\circ}$, they result in a more pronounced $\hat{\theta}$ component to $g$. This is predominantly in the $+ \hat{\theta}$ direction, and thus poleward.

Figure \ref{fig1} expands our view to the full set of spatial variations of $g$ by plotting $g$ as a function of latitude and longitude. Overall, Haumea's surface gravitational acceleration varies by almost two orders of magnitude---from 1.076 m/s$^2$ at the pole, to a minimum of 0.0126 m/s$^2$ at equatorial longitudes of 0 and 180$^{\circ}$ (for the rest of the manuscript, $g$ refers to the magnitude and is always positive). Along longitudes of -90 and 90$^{\circ}$ (corresponding to the minor), $g$ at the equator (0.20 m/s$^2$) is a factor of 5 lower than at the pole. On Haumea, along the -90 and 90$^{\circ}$ longitude meridians, $g$ is at its maximum for a given latitude. For context, at Earth, $g$ is only 0.5\% stronger at the poles than at the equator, and there are no longitudinal variations.

\section{Crater dimensions}
\subsection{Methods: Crater volumes}
\label{prop:mthd:vol}

Our computation of Haumea's surface gravity field enables calculation of key crater properties across its surface. Perhaps the most fundamental cratering property of interest to predict is the crater volume that would be expected for an impactor of a given size. We begin our calculations with the determination of crater volumes as a function of impactor and target body properties, and then expand our analysis to calculate depths and diameters using observationally derived depth to diameter ratios. To make predictions for crater volumes on Haumea, we use the scaling methods developed over thirty years in the works of \citealt{holsapple1982}, \citealt{housen1983}, \citealt{schmidt1987}, \citealt{holsapple1993}, and \citealt{holsapple2012}, including other references therein. These crater scaling laws approximate the impact as a point source and use dimensional analysis to relate the properties of the impactor and target to the characteristics of the resulting crater. The predicted crater properties include morphology (simple vs complex), volume, ejecta velocity and thickness . We will refer to these relations throughout the manuscript as the cratering ``point-source solutions.''

We have adopted the point-source solutions due to the availability in the literature of experimentally derived scaling constants for this formulation relevant to impacts in ice. The point-source solutions have compared favorably with other methods of predicting crater diameters. Although in application to the Earth, \citealt{johnson2016} find only 10\% discrepancies in the crater diameters predicted by \citealt{holsapple2012} with a numerical model, as well as three other scaling laws. \citealt{kurosawa2019} derive an alternative, fully analytical (i.e. without use of dimensional analysis) formulation for predicting crater diameters and find their range of predicted crater diameters to be wholly consistent with those of \citealt{schmidt1987}. One of their ejecta velocity distribution solutions matches those of \citealt{housen1983}, although two other solutions can vary by up to an order of magnitude.

The cratering point-source solutions are defined in two limits, based on the relative material strength of the planetary surface compared to the lithostatic pressure. In the ``strength regime,'' the crustal strength is large compared to lithostatic pressure. Conversely, in the ``gravity regime'' instead of target strength, the weight of the excavated material is the principal force preventing crater growth. \citep{holsapple1993}. The transition from strength to gravity regime is such that they correspond to smaller and larger craters respectively.

\citealt{holsapple1993} derive a relation (equation 18 of that work, equation \ref{eq_vol} below here.) for the non-dimensional cratering efficiency $\pi_v$, which relates the ratio of the mass of material originally contained within the crater profile, the crater mass, and the mass of the impactor. This is frequently referred to as the pi-group scaling.

\begin{align}
\begin{split}
  \label{eq_vol}
  & \pi_v = K_1 \biggl\{ \pi_2 \biggl( \frac{\rho}{\delta} \biggr) ^{(6 \nu - 2 - \mu)/3 \mu} \\
          & \hspace{45pt} + \biggl[ K_2 \pi_3 \biggl( \frac{\rho}{\delta} \biggr)
          ^{(6 \nu - 2 )/3 \mu} \biggr]
          ^{(2 + \mu)/2} \biggr\} ^{-3\mu/(2 + \mu)}
\end{split}
\end{align}

\begin{align}
\begin{split}
  \label{pi_group}
    \pi_v = \frac{\rho V}{m}, \qquad \pi_2 = \frac{g a_i}{U^2}, \qquad \pi_3 = \frac{Y}{\rho U^2}
\end{split}
\end{align}

Here the target body properties are density $\rho$, local surface gravity $g$, and material cohesive strength $Y$ (in dimensions of stress). Crater volume $V$ is the result of target body properties. The impactor properties are radius $a_i$ (thus assuming a spherical body), impact velocity $U$, and mass $m$ (where $m$ = (4/3) $\pi \delta a_i^3$.). $\pi_v$, $\pi_2$, and $\pi_3$ are non-dimensional parameters for the cratering efficiency, gravity-scaled size, and strength respectively.

\begin{figure}
  \centering
  \epsscale{1.2}
  \plotone{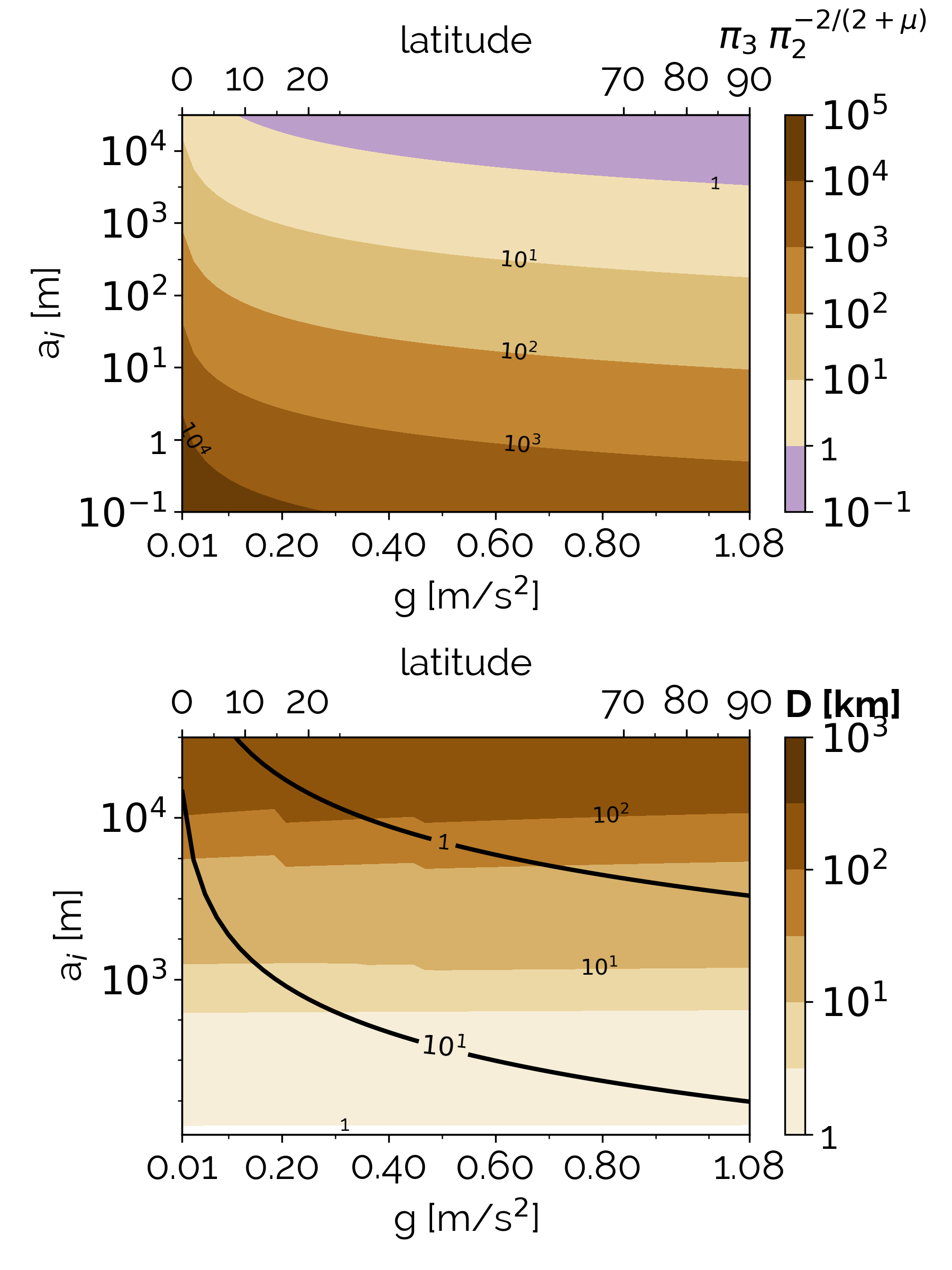}
  \caption{(a) Contours of the ratio of surface strength to gravitational forces ($\pi_2 \pi_3 ^{-2/(2+\mu)}$) as a function of surface gravity ($g$) and impactor radius ($a_i$), from which the strength to gravity regime transition can be interpreted. Specifically, the regime transition occurs when 0.1 $<$ $\pi_2 \pi_3 ^{-2/(2+\mu)}$ $<$ 10. A second x-axis on the top shows the latitudes corresponding to the $g$ values on the bottom, wherein unique latitudes corresponding to 0.27 $<$ $g$ $<$ 0.8 m/s$^2$ cannot be labeled due to the existence of multiple latitudes that correspond to each of these $g$ values. (b) The shaded contours here show crater diameters ($D$) that result as a function of surface gravity ($g$) and impactor radius ($a_i$). The curves are for the same $\pi_2 \pi_3 ^{-2/(2+\mu)}$ parameter shown as contours in subplot a). This allows for reading off of the crater diameters that correspond to the strength to gravity regime transition.}
  \label{fig3}
\end{figure}

The constants $K_1$ and $K_2$, as well as exponents $\mu$ and $\nu$ are fit to from experimental data. $K_2$ is commonly set to 1, as we do here, such that $K_1$ as well as the exponents $\mu$ and $\nu$ are determined from experiments with specific materials. These constants encapsulate the energy and momentum coupling between the impactor and surface, Experiments must be at field, not lab, scale for relevance to the large craters visible at the resolutions of spacecraft observations (specifically spanning the strength and gravity regimes). Such field experiments cannot be carried out at the pressure and temperature conditions of the outer solar system. Thus, the best available option is to adopt for numerical constants, the values informed from terrestrial field observations of explosive craters in ice, as is done in \citealt{holsapple2022}. The craters from these observations occur in the strength regime. We have, however, the benefit of being able to tune our predictions to remote sensing observations of icy bodies such as the Saturnian satellites. We find that the cold ice value for $K_1$, specifically, predicts crater diameters an order of magnitude too large compared with what is observed on the Saturnian satellites. This is likely related to the greater material strength of ice at outer solar system conditions compared to the coldest practicable conditions on Earth. For this reason, we adopt the $K_1$ for hard rock as opposed to cold ice (see discussion at end of section \ref{prop:mthd:trans}).

The relative magnitudes of the gravity-scaled size ($\pi_2$) and strength group ($\pi_3$), defined in equation \ref{eq_vol}, specify whether cratering is occurring in the strength or gravity regimes. Specifically, per \citealt{holsapple1987}, the strength to gravity transition is found to occur when:

\begin{align}
\begin{split}
  \label{eq_transn}
  0.1 < \pi_3 \pi_2 ^{-2/(2+\mu)} < 10
\end{split}
\end{align}

\subsection{Methods: Simple to complex transition}
\label{prop:mthd:trans}.

The two most numerous types of crater morphology are simple and complex. Simple craters are bowl-shaped---despite some wall collapse and downslope deposition, the transient crater from the impact event is largely preserved. Complex craters are generally considered to be the result of the gravity-driven collapse generating upward and inward flow, manifesting in terraced walls, central peaks and flat floors. The net effect is a depth to diameter ratio that is smaller than for simple craters, although this ratio varies as a function of crater size \citep{pike1977}. Simple craters occur in both the strength and gravity regimes of the point-source solutions, while complex craters are only found in the gravity regime \citep{holsapple1993}.

We use the point-source analytical relations to predict crater volumes and partially solve for crater diameters for a given simple or complex crater. The simple to complex transition can be roughly predicted with equation \ref{eq_transn}. Nevertheless, to predict a complete morphometry, including both depth and diameter, as well as a prediction of simple or complex morphology we use constraints from spacecraft observations of cratering into icy bodies. This sample has benefitted in recent years from observations by the \textit{Cassini} spacecraft of the Saturnian satellites as well as the \textit{New Horizons} mission's observations of Pluto and its largest satellite Charon. Specifically, we use observationally derived fits to the simple crater to complex crater transition diameter ($D_t$) as a function of surface gravity (discussed in this section), as well as the complex crater depth to diameter (d/D) ratio as a function of gravity (discussed in section \ref{prop:mthd:dim}). Studies spanning two decades have quantified the simple to complex transition across icy bodies imaged at high resolutions \citep{schenk1989,schenk1991,moore2001,schenk2002a,white2017,robbins2021,schenk2021}.

\citealt{aponte-hernandez2021} looked at the simple to complex transition diameter ($D_t$) for Saturn's satellite Rhea, but most relevant to this work derive a fit for $D_t$ as a function of surface gravity for a total of 16 icy bodies. These include the major icy satellites of the giant planets, as well as Pluto, Charon and Ceres. The measurements for the 15 bodies asides from Rhea are the result of other studies, and we refer the reader to Figure 11 of \citealt{aponte-hernandez2021} for the references of measurements of individual bodies. They find that $D_t$($g$) is described by a power law, with the specific fit for icy bodies being: $D_t$ = (39.7 $\pm$ 1.7 km)$g^{-0.4 \pm 0.1}$, with $g$ here in units of cm/s$^2$ (all other relations use SI units unless otherwise noted). The data used to create the power law fit span Mimas (0.064 m/s$^2$) -- Ganymede (1.428 m/s$^2$), which encompasses part of the range of estimated gravities for Haumea (0.0126 -- 1.076 m/s$^2$). However, the lowest predicted gravities for Haumea are ~5x smaller than the smallest icy body with a documented simple to complex transition.

It is in comparison with the observed crater diameters corresponding to the simple to complex transition for Pluto and Charon \citep{robbins2021}, as well as the Saturnian satellites Dione and Tethys \citep{white2017}, that we find an order of magnitude too large crater diameters if we adopt the cold ice value for constant $K_1$. The $K_1$ value for hard rock results in predictions with the correct of order magnitude. For this comparison, we calculate crater diameters from volumes as delineated in section \ref{prop:mthd:dim}.

\begin{figure}
  \centering
  \epsscale{1.2}
  \plotone{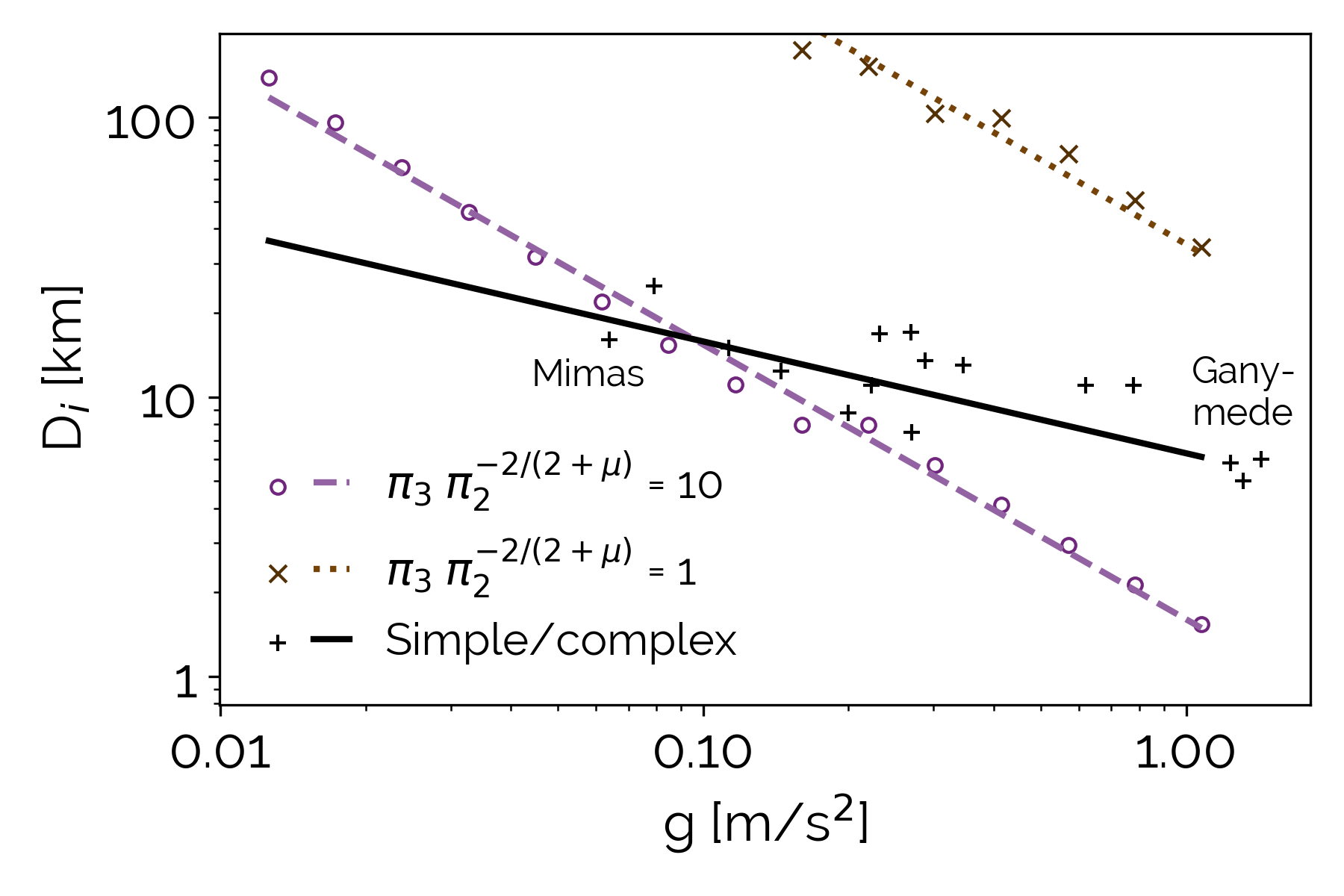}
  \caption{The power law simple to complex crater transition for icy bodies as a function of gravity, as calculated by Apont\'{e} Hernandez et al. 2021, is shown as the black line. The points for the planetary bodies that that work fit to are in order of increasing $g$: Mimas, Miranda, Enceladus, Tethys, Umbriel, Iapetus, Dione, Ariel, Ceres, Charon, Oberon, Pluto, Triton, Callisto, Europa, and Ganymede. We show when the ratio of surface strength to gravitational forces ($\pi_2$ $\pi_3 ^{-2/(2+\mu)}$) for cold ice are equal to 10 in purple, and to 1 in brown. These are calculated numercially at the locations of the circles or X's, to which the dashed and dotted lines are fitted.}
  \label{fig4}
\end{figure}

\subsection{Methods: Crater dimensions}
\label{prop:mthd:dim}

\begin{deluxetable*}{ccccccc}
\tablehead{
\colhead{Bin \#} & \colhead{Bin Gravity (m/s$^2$)} & \colhead{Planetary Body} & \colhead{Actual Gravity (m/s$^2$)} & \colhead{$\alpha$} & \colhead{$\beta$} & \colhead{n}
}
\decimalcolnumbers
\startdata
Simple    & 0.012 -- 1.08 & Tethys                       & 0.147                      & 0.299 & 0.832 & 55 \\
Complex 1 & 0.012 -- 0.2  & Tethys                       & 0.147                      & 0.458 & 0.662 & 17 \\
Complex 2 & 0.2 -- 0.45   & Iapetus, Dione, Rhea, Charon & 0.223, 0.233, 0.264, 0.288 & 0.446 & 0.544 & 67, 38, 48, 46 \\
Complex 3 & 0.45 -- 1.08  & Pluto                        & 0.62                       & 0.346 & 0.546 & 60 \\
\enddata
\caption{Power law coefficients for crater depth to diameter ratios in the adopted surface gravity bins. The power law fits for the Saturnian satellites are from \citealt{white2017}, while those for Pluto and Charon are from \citealt{robbins2021}.}
\label{table2}
\end{deluxetable*}

In calculating crater dimensions from crater volumes, specifically depth and diameter, we first distinguish between simple and complex craters. Simple craters show depth to diamter ratios that are largely independent of surface gravity. Simple crater depth to diameter ratios average to roughly 0.2 across all planetary bodies \citep{pike1977,robbins2018a}. Nevertheless, notable variability is found for icy bodies, which may represent different methodologies, sample size issues, or an effect of material properties (see Figure 6 of \citealt{robbins2018a}, Figure 3 of \citealt{white2017}).

In order to back out the diameter ($D$) for a simple crater that would correspond to a calculated crater volume ($V$), we use the  following relation from \citealt{holsapple2022}:

\begin{align}
\label{simple_D}
  & D = 2 K_r V^{1/3}
\end{align}

with the values for constant $K_r$ taken from data and suggested to be equivalent for all cohesive materials (the best available data for ice are explosive craters at terrestrial conditions). Specifically, $K_r$ = 1.1.

To calculate simple crater depths, we use the depth to diameter ratio that has been observed for Tethys \citep{white2017}. This is because for the crater volumes that we investigate in section \ref{res:dim}, simple craters are only expected to form at the low end of Haumea's surface gravity range and thus transition well to our lowest complex depth to diameter ratio bin, which is also from Tethys (Table \ref{table2}). Tethys is an icy body with relatively low ($<$ 0.2 m/s$^2$) surface gravity, that also has a large statistical sample of crater depth to diameter ratios. Other icy bodies with $g$ $<$ 0.2 m/s$^2$ have either small statistical samples (Mimas) or considerable internal heat resulting in crater relaxation (Enceladus).

For complex craters, the depth to diameter ratio is not constant, and follows a power law dependency \citep{pike1977,holsapple1993,white2017,robbins2021}:

\begin{align}
\label{cmplx_d}
  d = \alpha D^{\beta}
\end{align}

where $d$ and $D$ are the crater depth and diameter, respectively, in units of km for the purposes of these observationally derived fits. The values of the exponents in the power law are found to be a function of surface gravity \citep{robbins2018a,white2017,aponte-hernandez2021}. Because of the large range in the surface gravity across Haumea's surface, for our calculations we need to account for the power law exponent varying as a function of surface gravity. To accomplish this, we use three surface gravity bins, obtained from fits to depth to diameter ratios on icy bodies, namely the Saturnian satellites \citep{white2017}, and Pluto and Charon \citep{robbins2021}. The surface gravity on these bodies ranges from 0.145 $<$ $g$ $<$ 0.62 m/s$^2$, and the fits for the bodies with the lowest (Tethys) and highest (Pluto) surface gravities are extrapolated to cover the full range of surface gravity on Haumea (0.0126 m/s$^2$ $<$ $g$ $<$ 1.08 m/s$^2$). The bins, and the specific fit parameters $\alpha$ and $\beta$ are shown in Table \ref{table2}.

For the other dimensions that form the shape of the complex crater, we assume a flat crater floor and uniform slope from the crater floor diameter ($D_f$) to the rim, or overall, crater diameter $D$. We choose a flat floor for computational simplicity, while some complex craters have central peaks, a paramterization for a peak shape would have to be adopted.

\begin{align}
\label{cmplx_V}
  & V = \frac{\pi d}{4} \left[D^2 + \frac{1}{3} (D - D_f)(D + 2 D_f)\right]
\end{align}

Where the flat floor diameter ($D_f$) is set equal to 0 at the transition diameter to simple craters ($D_t$), and related to the overall crater diameter $D$ using fits to lunar crater profiles \citep{pike1977,holsapple2022}. An equivalent parametrization for icy crater $D_f$ is not available in the literature.

\begin{align}
\label{cmplx_Df}
  D_f = 0.292 (2 D_t) ^{-0.249} (D - D_t) ^{1.249}
\end{align}

We are now able to solve for the crater morphometry that would correspond to an impactor of a given size anywhere on Haumea's surface. Equations \ref{cmplx_d} -- \ref{cmplx_Df} are solved simultaneously to determine the complex crater diameter $D$ and depth $d$ that correspond to a crater of volume V. Specifically, Equation \ref{eq_vol} is used to determine the crater volume corresponding to an impactor of given radius. Equation \ref{simple_D} is then used to calculate the diameter of a simple crater corresponding to this volume. If the diameter is less than the simple to complex transition diameter, Equation \ref{cmplx_d} can be applied with the simple crater coefficients from Table \ref{table2} to calculate a depth. If the calculated simple crater diameter is greater than the simple to complex transition, then Equations \ref{cmplx_d} -- \ref{cmplx_Df} are solved numerically to obtain the unique $d$ and $D$ that correspond to the known $V$.

\subsection{Results: Crater transitions}
\label{prop:res:dim}

For Haumea, the transition between the strength and gravity regime for cratering begins begins for an impactor radius of 200 m at the pole, compared to a radius of 15,000 m at the equator (Figure \ref{fig3}a). These quoted values are for gravity-scaled size to strength ratios ($\pi_3 \pi_2 ^{-2/(2+\mu)}$) of 10. These impactor radii for the strength to gravity regime transitions correspond to crater diameters of 0.5 and 150 km respectively (Figure \ref{fig3}b).

To visualize the crater diameters at which crater morphologies transition from simple to complex, we plot the data from multiple works tabulated in \citealt{aponte-hernandez2021}, as well as the \citealt{aponte-hernandez2021} power law fit to these data, as the black crosses and line respectively in Figure \ref{fig4}. These are plotted over the range of surface gravity found on Haumea. Crater diameters below the \citealt{aponte-hernandez2021} relation at a given surface gravity are expected to be simple, while above the line complex craters are expected. Also plotted are lines for various values of $\pi_3 \pi_2 ^{-2/(2+\mu)}$, indicating the strength to gravity regime transition, and thus the theoretical prediction of the parameter space in which complex craters should begin appearing. Specifically, complex craters should begin appearing above the $\pi_3 \pi_2 ^{-2/(2+\mu)}$ = 10 line. For $g$ $>$ 0.1 m/s$^2$, complex craters only occur after the beginning of the strength to gravity regime transition, as would be expected. However, for $g$ $<$ 0.1 m/s$^2$, complex craters are suggested to occur partly in the strength regime, indicating that improvements in our understanding of crater transitions in icy bodies for $g$ $<$ 0.1 m/s$^2$ are necessary (see Discussion). We remind the reader that we have selected our value of constant $K_2$ based on comparison with crater diameters used for four bodies in the \citealt{aponte-hernandez2021} relation. Thus it is not surprising that our prediction for the strength to gravity regime transition matches the simple to complex transition around the surface gravities of these mid-sized bodies (Tethys, Dione, Charon, Pluto). This involved, however, simply adopting the \citealt{holsapple2022} $K_1$ value for hard rock rather than cold ice. That the slope of our regime transition curves, and the \citealt{aponte-hernandez2021} simple to complex transition differ suggest that additional studies of an appropriate value for this constant would be beneficial.

\subsection{Results: Crater volume and dimensions}
\label{res:dim}

We first investigate how the crater volume varies as a function of gravity, impactor velocity, and size, to understand sensitivities in the parameter space and to focus our further modeling efforts.

\begin{figure}
  \centering
  \epsscale{1.2}
  \plotone{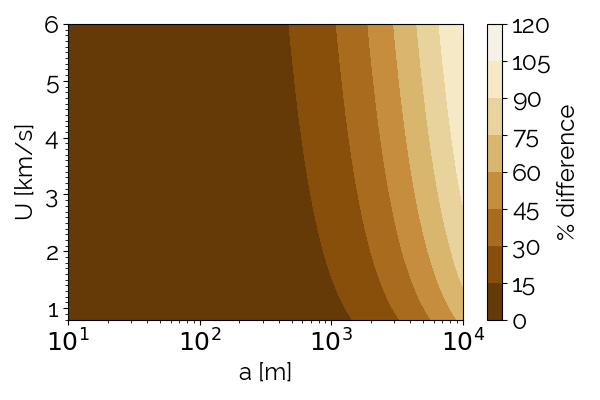}
  \caption{The percent difference between crater volumes ($\Delta$V) at Haumea's maximum (polar) and minimum (equatorial) surface gravities, as a function of impactor velocity ($U$) and radius ($a_i$).}
  \label{fig5}
\end{figure}

In Figure \ref{fig5} we plot the predicted percent difference in crater volumes at the area of maximum (the poles, g = 1.08 m/s$^2$) and minimum (location of equatorial major axis, g = 0.0126 m/s$^2$) surface gravity, as a function of impactor velocity ($U$) and radius ($a_i$). The 1 -- 6 km/s range for $U$ is taken from statistical studies of Kuiper Belt Object impacts \citep{delloro2013,greenstreet2015}. These studies find the mean velocity to be 2 -- 2.5 km/s. The general trend is for greater percent differences in volume ($\Delta$V) for increasing impactor radius and impact velocity. $\Delta$V remains under 30\% for $a_i$ $<$ 10$^3$ m, regardless of impact velocity. The percentage difference in expected volumes changes much more with respect to variations in impactor size rather than impactor velocity, as shown by the steepness of the contours. For an impactor with a = 10$^4$ m, the percent difference in the volume of the resultant crater at Haumea's pole vs equatorial major axis can exceed 100 \%.

From these results, we focus on quantifying variations as a function of impactor radius for the rest of the manuscript, and for all later calculations in the manuscript which require specifying an impact velocity, we use $U$ = 2 km/s, the approximate average impact velocity expected for Kuiper Belt Objects \citep{delloro2013}. Within the impactor radius parameter space, we focus on 0.5 $<$ $a_i$ $<$ 16 km. The limit on the low end is due to the small effect that the surface gravity has on crater volume at impactor sizes of $<$ 1 km, as described above. On the upper end, a $\sim$15 km impactor results in craters of $\sim$150 km. 150 km is the largest diameter crater that could be expected to be found on Pluto were its surface age $>$ 1 Gyr \citep{greenstreet2015}, which we consider a reasonable upper bound to consider.

\begin{figure}
  \centering
  \epsscale{1.2}
  \plotone{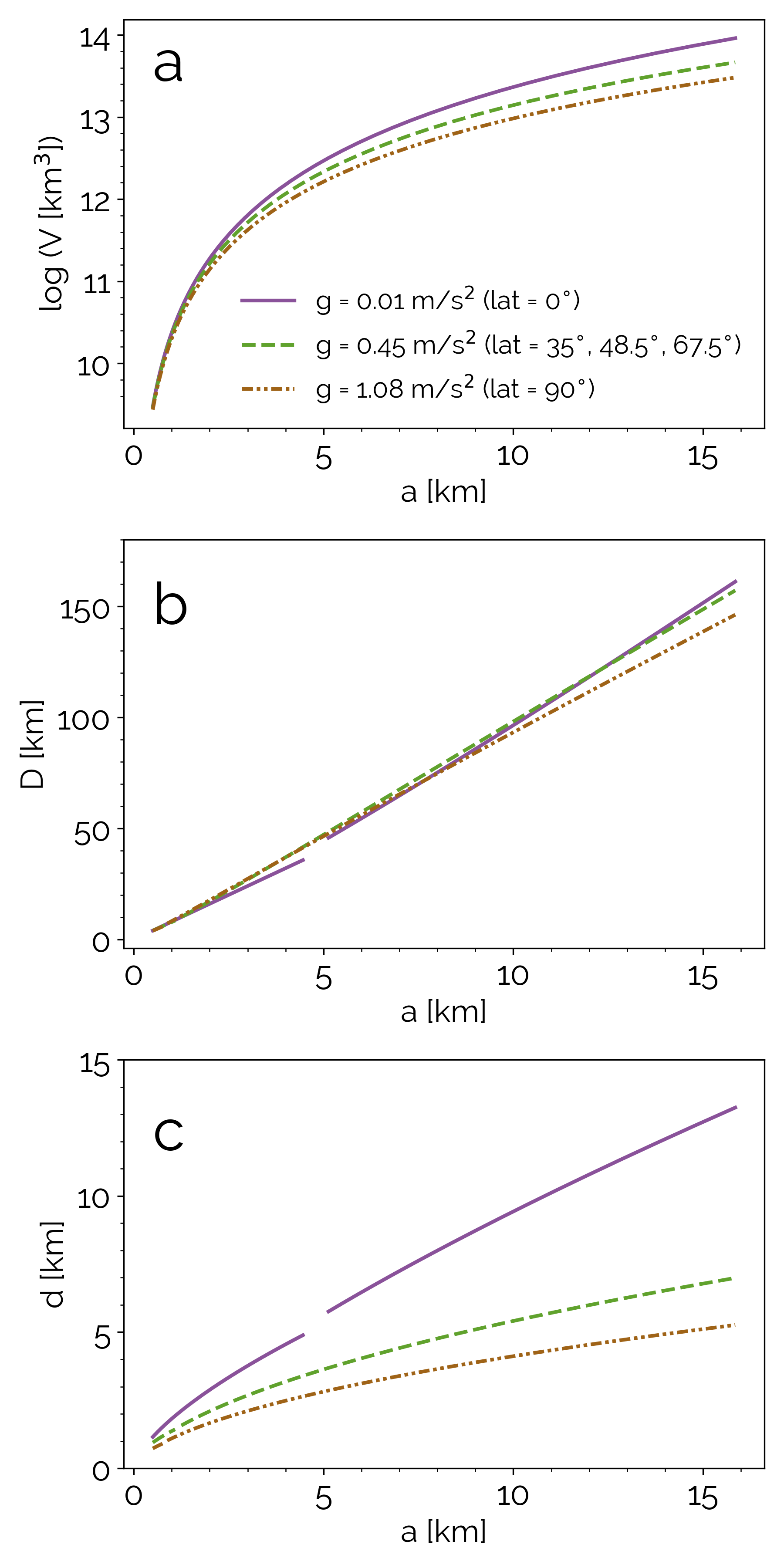}
  \caption{(a) The logarithm of the crater volumes ($V$) as a function of impactor size for the surface locations corresponding to 3 different surface gravities. (b) The crater diameters ($D$) for the craters whose volumes are shown in subplot a). (c) The crater depths ($d$) for craters whose volumes are shown in subplot a).}
  \label{fig6}
\end{figure}

In Figure \ref{fig6}, we plot crater volumes, diameters ($D$), and depths ($d$) at 3 representative surface gravities (in turn, representing 5 latitudes at longitude = 0$^{\circ}$) on Haumea for the aforementioned range of impactor radii 0.5 $<$ $a_i$ $<$ 16 km. Recall that per section \ref{gravity:rslt}, multiple latitudes in the same hemisphere can have the same $g$ value. These surface gravities are specifically selected to span the 3 different gravity bins used in calculating the crater $d/D$ ratio tabulated in Table \ref{table2}.  Crater volumes start diverging appreciably as a function of surface location for impactors $a_i$ $>$ 2 km (Figure \ref{fig6}a). Comparing the calculated crater diameters (Figure \ref{fig6}b) and depths (Figure \ref{fig6}c), it becomes apparent that most of the difference in crater volume is accomodated by variations in depth rather than diameter. Crater diameters are largely consistent among the different locations on Haumea until impactors reach radii of $a_i$ $>$ 10 km. Even then, the differences in diameter are consistently $<$ 20 \%, while differences in depth can exceed 300\%, with a 47\% difference even between the more similarly scaled polar and mid-latitudes.

From the \citealt{aponte-hernandez2021} relation (Figure \ref{fig4}), we infer a simple to complex transition diameter ($D_t$) of 36.2 km at Haumea's location of minimum surface gravity at the equator, compared to $D_t$ = 6.1 km at the poles (Figure \ref{fig4}). Because the transition occurs at $D$ $<$ 10 km for $g$ $>$ 0.4 m/s$^2$, the transition is barely visible for the two higher surface gravity bins in Figure \ref{fig6}b, c. Nevertheless, the transition occurs at close to 30 km at the equator, with the transition visible as the break in the g = 0.01 m/s$^2$ line in the depth and diameter plots. The close match across the simple to complex transition for g = 0.01 m/s$^2$ is the result of our using observationally derived data for the same planetary body, Tethys, for both simple and complex craters at this surface gravity (Table \ref{table2}, first two rows).

\section{Ejecta thickness}
\subsection{Methods: Ejecta thickness}
\label{ej:mthd:reg}

In investigating how crater ejecta thickness varies across Haumea's surface, and as a function of its variable surface graviy, we note that the point-source solutions only predict surface gravity to affect ejecta thickness in the strength regime \citep{housen1983}. This is a result of the strength ($Y$) and $g$ being grouped into the same term $Y$/$\rho g R$ from dimensional analysis. This ratio becomes very small in the gravity regime (see \citealt{housen1983}, equations 32 and 33), and as this term is the only one containing $g$ for relations related to ejecta thickness, the effect of $g$ is negligible in the gravity regime.

In the strength regime, \citealt{housen1983} finds that the ejecta thickness can be calculated as:

\begin{align}
  \begin{split}
    \label{eq_thick_full}
    \frac{B(r)}{R} = \frac{A (e_r - 2)}{2 \pi} & (\sin 2 \theta)^{e_r - 2}
    \biggl(\frac{r}{R}\biggr)^{-e_r} \\
    & \times \biggl[ 1 + \frac{4e_r - 5}{3}
    \biggl( \frac{r}{R \sin 2\theta}^{-(e_r - 2)/2} \frac{D}{r} \biggr) \biggr]
  \end{split}
\end{align}

with $B(r)$ being the ejecta thickness as a function of radial coordinate $r$, $R$ the crater radius, and $\theta$ the impact angle with respect to the horizontal. This relation applies for $r$ $>$ $R$. The exponent $e_r$ is defined as:

\begin{align}
  e_r = \frac{6+\alpha_E}{3-\alpha_E}
\end{align}

and constant $A$ is defined in the strength regime as:

\begin{align}
  \begin{split}
    A = K_4 \left( \frac{Y}{\rho g R} \right)^{3\alpha_E/(3 - \alpha_E)}
  \end{split}
\end{align}

with the exponent $\alpha_E$ being related to the exponent $\mu$ used throughout the manuscript:

\begin{align}
\label{eqn_alpha}
  \alpha_E = \frac{3 \mu}{2+\mu}
\end{align}

As with the exponent $\mu$, $\alpha_E$ is ultimately encapsulating energy and momentum transfer between the impactor and surface. However, the form of $\alpha_E$ specifically captures how the cratering efficiency ($\pi_V$ from Equation \ref{eq_vol}) decreases with increasing impactor size in the gravity regime. $D$ in equation \ref{eq_thick_full} is defined in the strength regime as:

\begin{align}
  \begin{split}
    D = \left( \frac{(K_2)^2 Y}{\rho g R} \right)^{-b}
  \end{split}
\end{align}

with exponent b defined as:

\begin{align}
  \begin{split}
    b = \frac{\alpha_E - 3}{4 \alpha_E}
  \end{split}
\end{align}

For the ejecta thickness calculations, we derive the value for $\alpha_E$ from the Eulerian shock physics numerical simulations of \citealt{senft2008}, who simulated the results of a 100 m basalt impactor into a 200 m thick ice layer on Mars. The value we derive of $\alpha_E$ = 0.6117 is comparable to the value of 0.6471 that would be calculated using equation \ref{eqn_alpha} from our adopted $\mu$ value of 0.55 (the latter motivated by terrestrial observations of explosive craters into ice).

Because the constant $K_4$ must be determined experimentally, and appropriate ejecta experiments or simulations into ice in the strength regime are not available (the \citealt{senft2008} Mars simulations lie in the gravity regime, which allows for determining $\alpha_E$ but not $K_4$) it is not possible to calculate actual ejecta thicknesses for Haumea. Rather, we calculate the relative thickness of the ejecta at all latitudes ($B_g$), compared to the thickness at the poles where gravity is at a maximum ($B_{g,max}$). Due to the inverse relation between $B$ and $g$, ejecta thickness would be at a minimum at the poles. By taking the ratio of equation \ref{eq_thick_full} for the ejecta thickness at a given latitude ($B_g$) to that at the equator ($B_{g,max}$), one can calculate the ejecta thickness relative to the equator:

\begin{align}
\label{eq_thick_ratio}
  \frac{B_g}{B_{g,max}} = \left( \frac{g_{max}^{3\alpha_E/(3 - \alpha_E)} + g_{max}^{3\alpha_E/(3 - \alpha_E) - b}}{g^{3\alpha_E/(3 - \alpha_E)} + g^{3\alpha_E/(3 - \alpha_E) - b}} \right)
\end{align}

\begin{figure}
  \centering
  \epsscale{1.2}
  \plotone{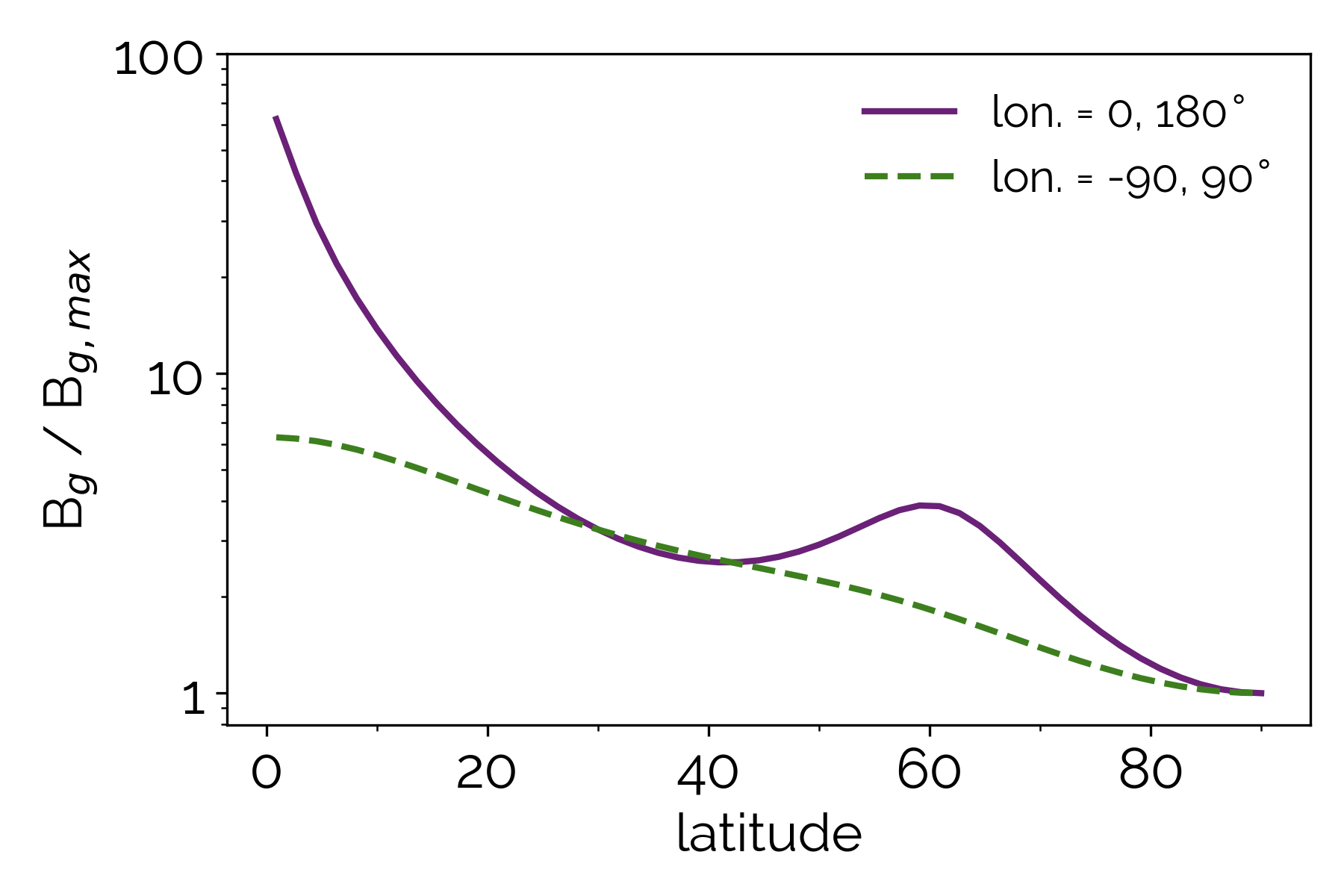}
  \caption{The ratio of the ejecta thickness ($B_g$) to the ejecta thickness at the location of Haumea's maximum surface gravity ($B_{g,max}$), as a function of latitude. $B_g$/$B_{g,max}$ is shown for longitudes of 0, 180$^{\circ}$ and -90, 90$^{\circ}$.}
  \label{fig7}
\end{figure}

\subsection{Results: Ejecta thickness}

As discussed in section \ref{ej:mthd:reg}, because the ejecta thickness does not vary as a function of surface gravity in the gravity regime, for craters larger than the strength-to-gravity regime transition size of $\sim$ 0.5 km at the pole and 150 km at the equator, impactors of the same size will result in ejecta of the same thickness regardless of location on Haumea.

For craters in the strength regime (smaller than the transition diameter, which varies from 100 m to 10 km depending on location on the surface, see section \ref{prop:res:dim}
), we plot the ejecta thickness ratio $B_g$/$B_{g,max}$ in Figure \ref{fig7}. $B_g$/$B_{g,max}$ is plotted for longitudes of 0, 180$^{\circ}$ and -90, 90$^{\circ}$. Because the longitudes of -90, 90$^{\circ}$ represent the meridians with the consistently highest surface gravity on Haumea, while 0, 180$^{\circ}$ are the meridians of minimum surface gravity, these two latitudes are the two end-members for $B_g$/$B_{g,max}$ as a function of latitude. $B_g$/$B_{g,max}$ for all other latitudes will fall between these two curves.

Moving equatorword from the pole, ejecta thicknesses at longitudes of 0, 180$^{\circ}$ quickly reach double their thickness at the pole, begininning at around $\sim$75$^{\circ}$ latitude. A local maximum with $\sim$4 times the thickness at the pole is oberved at 60$^{\circ}$, the same latitude at which a local maximum in $g$ is observed for these longitudes (Figure \ref{fig1}). For both longitude end members, ejecta thicknesses largely remain within a factor of 10 times the thickness at the pole. The exception is below 14$^{\circ}$ latitude at 0, 180$^{\circ}$ longitude where ejeca thicknesses rapidly increase as $g$ is approaching its minimum of 0.0126 m/s$^2$, ultimately reaching at the equator a factor of 63 times thicker than at the poles.

\section{Escape of ejecta}
The amount of escaping material will vary as a function of location on Haumea's surface, in the case of an impact large enough to eject material above the local escape velocity. This is a result of the spatial variations in surface gravity, coupled with the latitudinal variation in the tangential velocity of the surface from Haumea's rotation. This will also be a result of the minimum velocity of ejecta on Haumea, which we will demonstrate experiences even greater spatial variation than the escape velocity.

\begin{figure*}
  \centering
  \epsscale{1.15}
  \plotone{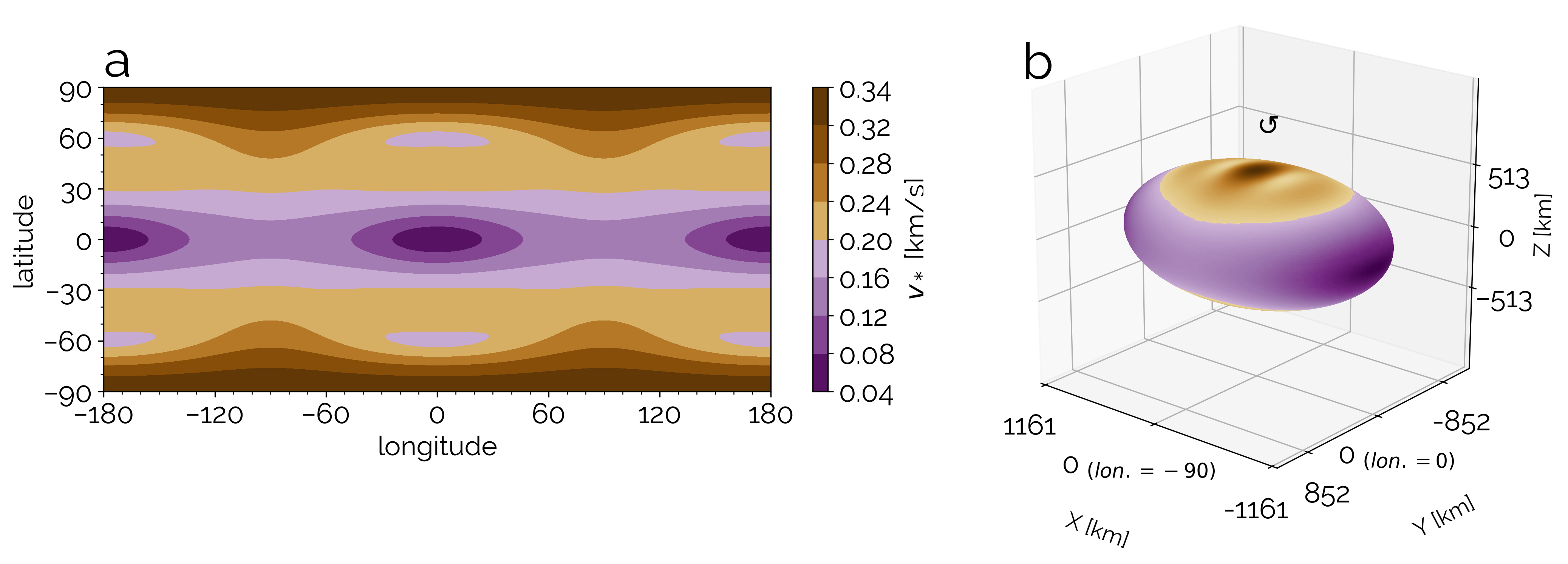}
  \caption{(a) The minimum ejecta velocity ($v_*$) as a function of latitude and longitude. (b) A 3D perspective of $v_*$. Plotting conventions are the same as in Figure 2.}
  \label{fig8}
\end{figure*}

\subsection{Methods: Escape velocity}

We first examine the ejecta velocities that would result for a large impact (i.e. in the gravity regime. Impactor larger than 200 m to 15 km, depending on location on the surface, see section \ref{prop:res:dim}).

The ejecta velocity ($v$) distribution exhibits a power law decay as a function of launch position, the power-law scaling is a fundamental result over the areas from which the influence of the impactor can be approximated as that of a point-source. The exceptions to this power law velocity distribution are in the immediate vicinity of the impactor, where the point-source approximation does not hold, and of the crater rim where gravity or strength stop the flow of ejecta. Thus very low velocity impacts, where the crater is only slightly larger than the impactor, would be the most notable scenario where a power law velocity distribution would not be representative of the majority of the ejecta \citep{housen2011,holsapple2012}.

Similarly, the mass fraction of ejecta ejected at velocities faster than velocity $v$ (which we represent as $M(v)$) is found to be a power law function of ejection velocity $v$ \citep{holsapple2012}. There is a minimum ejecta velocity $v_*$ for which this power law distribution applies. The ejecta below this velocity are the small mass that are ejected at the rim. The minimum ejecta velocity $v_*$ above which this power law distribution applies, can be calculated in the gravity regime as:

\begin{align}
\label{eq_v_star}
  v_* = K_{vg} \sqrt{ga}
\end{align}

With constant $K_{vg}$ derived from experiments directly measuring ejecta velocity distirbutions. We adopt the value of 3.3 derived for dense sand in \citealt{holsapple2012}. Dense sand is the most cohesive material for which experiments are available in the gravity regime. Other materials with greater target strength require experiments at proportionally larger (and experimentally impractical) scales to occur in the gravity regime. Because the gravity regime is specifically where material weight dominates over cohesion, we do not expect $K_{vg}$ to differ considerably for ice. In an ideal representation of $M(v)$ vs $v$, the two for all $v$ $>$ $v_*$ can be related as:

\begin{align}
\label{mass_frac}
  M(v) = M_e \left( \frac{v}{v_*} \right)^{-3 \mu}
\end{align}

where $M_e$ is the total ejecta mass, and in this idealized representation, is all ejected at or above the velocity $v_*$.

The other contributing factor to spatial variabilities in $M_{esc}$ will be variations in the local escape velocity ($v_{esc}$). The spatial variations in $v_{esc}$ are a result of both the variations in surface gravity, as well as the tangential velocity of the surface stemming from Haumea's rotation. We adopt a formulation that accounts for both of these effects, while making the simplified assumption of ejecta trajectories normal to the local surface \citep{scheeres1996}. This allows one to treat all ejecta trajectories equally. This avoids the need to consider at each point on Haumea's surface, how trajectory variations add or subtract to the local escape velocity. We calculate:

\begin{figure*}
  \centering
  \epsscale{1.15}
  \plotone{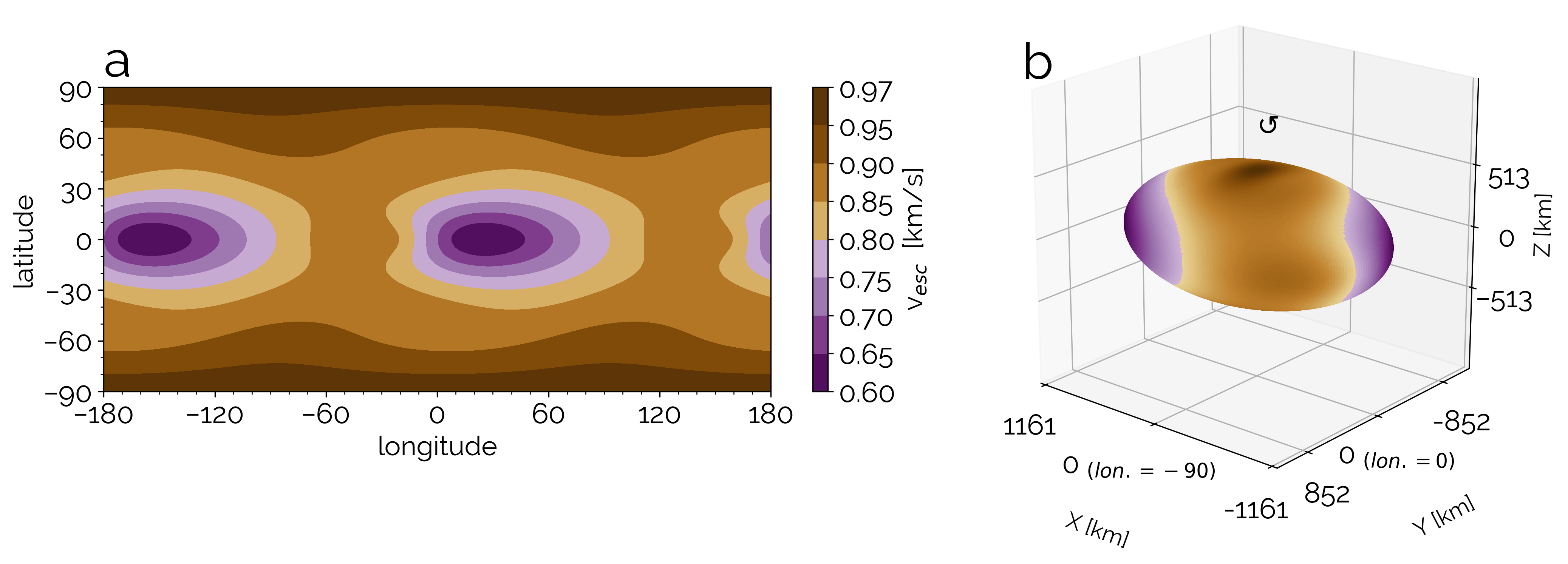}
  \caption{(a) The escape velocity ($v_{esc}$) as a function of latitude and longitude. (b) A 3D perspective of $v_{esc}$. Plotting conventions are the same as in Figure 2.}
  \label{fig9}
\end{figure*}

\begin{align}
\label{v_esc}
  v_{esc} = -\hat{n} \cdot (\Omega \times \vec{r}) +
              \sqrt{[\hat{n} \cdot (\Omega \times \vec{r})^2
              +2 U_{max} - (\Omega \times \vec{r})^2}
\end{align}

where $\hat{n}$ is the local surface normal, $\Omega$ is Haumea's angular velocity vector, and $\vec{r}$ is the vector from the origin to Haumea's surface. The quantity $U_{max}$ is a condition for particle escape, accounting for local variations in the gravity field and is calculated as:

\begin{align}
\label{U_max}
  \sqrt{2 U_{max}} = \text{max}[\sqrt{2 U(\vec{r})},\sqrt{2GM |\vec{r}|}]
\end{align}

with $G$ being the gravitational constant and $M$, Haumea's mass. $U(\vec{r})$ is the gravitational potential at the location $\vec{r}$ on Haumea's surface.

For our exploration of minimum ejecta velocities, $v_*$, as well as the fraction of total mass ejected, $M_{esc}/M_e$ (equation \ref{mass_frac} with $v_{esc}$ used for $v$) across Haumea's surface, we consider an impactor with radius ($a$) of 10 km. This represents a diameter for which appreciable differences in crater dimensions are seen across Haumea's surface (Figure \ref{fig6}).

\subsection{Results: Escape velocity}

In Figure \ref{fig8}, we plot the minimum ejecta velocities ($v_*$) for an impact of $a_i$ = 10 km. Because $v_* \propto \sqrt{g}$, the spatial variability in $v_*$ is similar to that in the surface gravity (Figure \ref{fig2}). The minimum in $v_*$ is at the equator at -155, 25$^{\circ}$ longitude, while being largest at the poles. Because of the square root on $g$, the variability in $v_*$ as a function of latitude is not as dramatic as $g$, increasing by a little over a factor of 8 at the poles compared to the equator.

The spatial variability of $v_{esc}$ is, in turn, plotted in Figure \ref{fig9}. $v_{esc}$ is higher at the poles compared to the equator, as is expected given that the surface gravity is stronger there. The latitudinal variation in $v_{esc}$ is 47\% from pole to the equator, substantially less than the latitudinal variability in both $g$, and $v_*$ for a 10 km impactor, although large in the context of other planetary bodies. Haumea also exhibits a relatively large longitudinal variability in $v_{esc}$ at the equator, at 38\%.

\begin{figure*}
  \centering
  \epsscale{1.15}
  \plotone{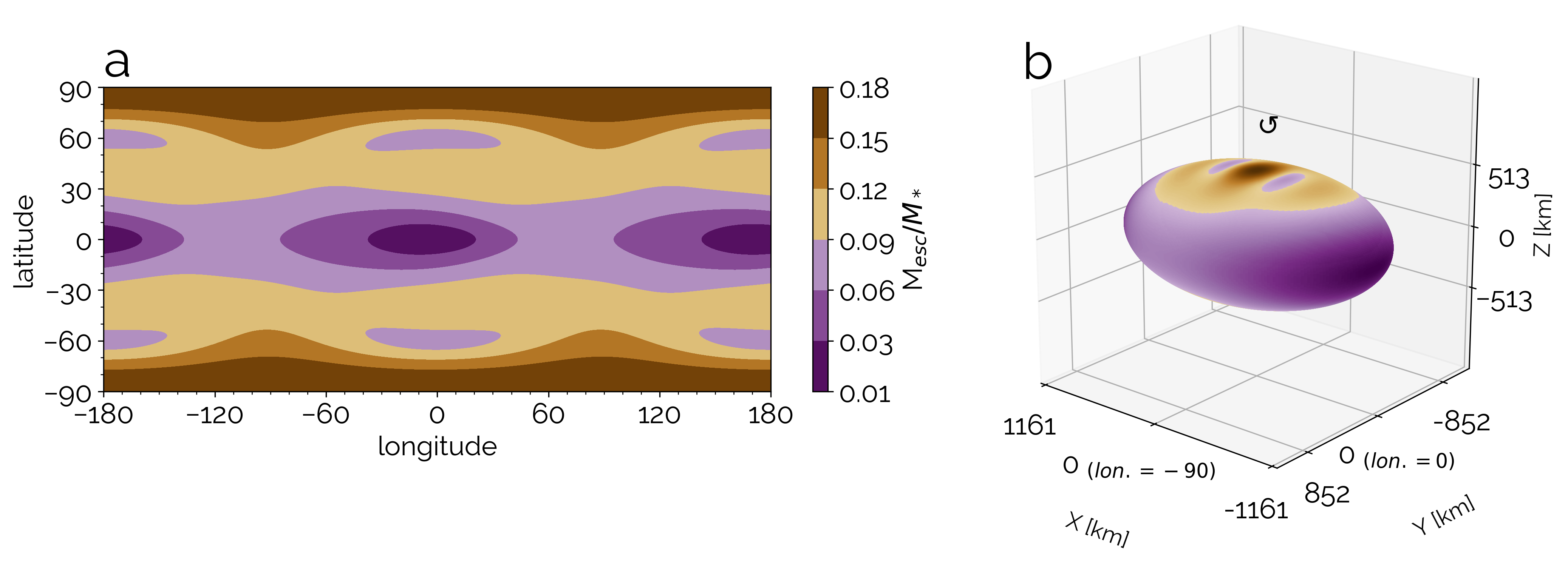}
  \caption{(a) The fraction of ejecta mass that escapes Haumea ($M_{esc}/M_*$) as a function of latitude and longitude. (b) A 3D perspective of $M_{esc}/M_*$. Plotting conventions are the same as in Figure 2.}
  \label{fig10}
\end{figure*}

In Figure \ref{fig10} we look at the spatial variability in the mass fraction of ejecta that escapes Haumea ($M_{esc}/M_e$) for an $a$ = 10 km impactor. The variability in $M_{esc}/M_e$ is a result of the spatial variations in both $v_*$ and $v_{esc}$. Despite Haumea's escape velocity being lower at the equator than at the poles, $M_{esc}/M_e$ is actually higher at the poles vs the equator---i.e. a greater fraction of ejecta escapes for an equivalent impactor at the pole. This is a result of the minimum ejection velocity $v_*$ showing greater variability from equator to pole (Figure \ref{fig8}) than $v_{esc}$ (Figure \ref{fig9}).

\section{Discussion}

That Haumea's centrifugal acceleration at the equator was comparable to its gravitational acceleration has been known previously; this is the explanations for Haumea's unique shape. However, focusing on impact processes, we have first explored here the implications for its surface environment, which are profound. Haumea's equatorial surface gravity at the locations of its major axis are almost two orders of magnitude lower than that at the poles. Furthermore, Haumea's large degree of flattening (i.e. a polar semi-major axis that is only 60\% the length of even the largest equatorial axis), results in surface normal vectors at higher latitudes ($>$ 60$^{\circ}$) that deviate greatly from being radially outward from Haumea's center of mass. This is manifested in strong $g_{\theta}$ gravitational terms, with Haumea's surface gravity vector at these latitudes pointing poleward relative to the surface normal. Finally, Haumea exhibits a non-monotonic increase in surface gravity strength with increasing latitude. This manifests in degeneracies in the latitudes with a given surface gravitational acceleration value between 25 and 70$^{\circ}$. While this is something that is seen on small bodies, it is certainly unique among known planet-sized bodies in the solar system.

For our calculations of the surface gravity strength, we assumed a uniform density for Haumea for ease of calculation of the spherical harmonic gravity coefficients. This is naturally an unrealistic assumption for Haumea, which is presumed to be differentiated \citep{dunham2019,noviello2022}. However, the primary factor resulting in Haumea's low equatorial surface gravity, which is the comparable magnitudes of the first order radial term ($g_{r,1}$) and the centrifugal acceleration ($\omega_r$), do not depend on this density assumption (see the expansion of these terms in the Appendix). A differentiated Haumea is expected to have a slightly smaller values of $g$ across the surface.

We predict Haumea's simple to complex transition diameter using observations of craters on 16 other icy bodies whose surface gravities span 95\% of Haumea's gravity range. In predicting the simple to complex transition diameter, we find that the transition diameter does largely occur outside of the strength regime as is predicted by the point-source relations. Nevertheless, below $g$ = 0.10 m/s$^2$, our extrapolation of observational data predicts the simple to complex transition occuring in the strength regime---a contradiction given that gravitational forces are ultimately responsible for the slumping that turns a transient crater into a complex crater. Improvements in our understanding of the tensile strength of ice at outer solar system temperatures and scales relevant for large impacts ($Y$) to refine our calculated strength-gravity regime transition would be helpful. Additional spacecraft observational studies of simple to complex crater transitions on icy bodies with $g$ $<$ 0.10 m/s$^2$ would be beneficial. With one exception, complex craters have seemingly not been observed on the Saturnian satellites Hyperion and Phoebe, but this would actually be more consistent with our calculated strength to gravity regime than from extrapolating the observationally derived trend from icy bodies (per Figure \ref{fig4}). While experimental studies of impacts into ice of sufficient scale are practically almost impossible, studies directly comparing scaling theory with the available data from explosion tests would be highly beneficial, although that is complicated by the relative inaccessibility of the military and company reports that contain these \citep{holsapple2022}.

For our examination of crater dimensions, we found that for impactors with radii $a$ $>$ 500 m, differences in the crater volume as a function of latitude will be acommodated by proportionally larger differences in crater depth than crater diameter. This is presumed to be a result of the high sensitivity of crater floor rebound during the formation process to surface gravity strength. The rebound alters crater depth more than diameter. Complex craters in environments with stronger surface gravity exhibit lower depth to diameter ratios than craters of equivalent volumes in lower surface gravity environments \citep{pike1977,holsapple1993,white2017,robbins2021}.

For craters in the strength regime (smaller than $\sim$0.5 km at the pole and 200 km at the equator), the range of variations in ejecta thickness is similar to that of the surface gravity, approaching two orders of magnitude. Higher surface gravity results in thinner ejecta deposits, because as gravity increases in the strength regime, ejecta deposits closer to the crater rim \citep{housen1983}. While we have focused on quantifying differences in relative ejecta thickness, the implication is that the radial extent of ejecta blankets at higher latitudes (with stronger surface gravity) will also be smaller compared to ejecta at lower latitudes. Our calculations suggest that the equatorial regions near Haumea's major axis will have ejecta blankets dramatically more noticeable than elsewhere on Haumea.

We note that for the spatial variations in the dimensions and ejecta thicknesses of Haumea's craters, we have focused on how crater properties will vary for an impactor of the \textit{same} size. In reality, Haumea will have been impacted by objects of many different sizes over its history. We emphasize that the effects that we have predicted will exhibit as statisical skews in crater characteristics as a function of location on Haumea's surface. Specifically, craters in Haumea's equatorial region will be preferentially deeper with thinner ejecta compared to craters in Haumea's polar regions. Haumea's mid-latitudes, which exhibit the degeneracy in surface gravitational acceleration as a function of latitude, will in turn exhibit a wider distribution of crater depths and ejeta thicknesses than exists in strictly the polar or equatorial regions.

The two order of magnitude variation in surface gravity across Haumea's surface, combined with Haumea's fast rotation rate, result in large variations in the escape velocity across Haumea's surface. Other planetary bodies in or near hyrdostatic equilibrium only exhibit latitudinal variations. Due to Haumea's shape, there also considerable longitudinal variations. Just along Haumea's equator, Haumea's escape velocity varies by 38\%.

We have demonstrated that Haumea's unique shape and short 3.92 hr day should manifest in dramatic expected variations in crater morphologies and morphometries across the same planetary body. The extent of these differences in crater types, volumes, depths, ejecta thicknesses, as well as ejecta retained during an impact are unique for currently known planet sized bodies in the solar system. The presence of crystalline ice on Haumea's surface provides a very rough potential constraint on the age of the surface. The calculation of crater characteristics in this work, could be combined with estimates of meteorite fluxes to create statistical predictions of the spatial distributions of crater types and sizes as a function of latitude and longitude.

There remain, also, numerous open areas of research in predicting peculiar characteristics of Haumea's environment as a result of its shape. With regards to the surface, tectonics as a result of the freezing out of a subsurface ocean is a strong possiblity, given predictions of Haumea having had such an ocean \citep{noviello2022}. While interior modeling suggests any differentiated core for Haumea would also be a triaxial ellipsoid, could the substantially lower equatorial gravity further concentrate communication outflows from the interior to the equatorial region? Haumea may thus be superficially similar to Charon, relatively cratered with equatorial tectonics. We have demonstrated, however, that the former will exhibit a much greater breadth of morphologies on Haumea than on Charon, with few complex craters and more noticeable ejecta at the equator, possibly similar to Mimas. The poles, in contrast, would have more complex craters, and the concentration of ejecta around the crater rim could manifest as the plateaus and other ejecta morphologies seen on Ganymede and Callisto.

Mass wasting may also be occurring, as impact shaking as has been proposed for the Saturnian satellites \citep{singer2012,fassett2014}. How this would be affected by a variable surface gravity field is an open question. Furthermore, how might Haumea's surface gravity variations affect its interior structure as well as its isostatic accomodation of surface features? Might the preferential escape of impact ejecta at polar vs equatorial latitudes have implications for the long-term evolution of Haumea's shape? Such questions warrant further investigation.

\section{Summary}

We have carried out the first detailed predictions of Haumea's surface morphology, focusing on the manifestation of craters. Given Haumea's surface composition being predominantly of inert water ice, we expect craters will be a dominant surface feature. We report the following findings:

\begin{enumerate}
  \item There is an almost two order of magnitude variation in Haumea's effective surface gravitational acceleration---from 1.076 m/s$^2$ at the pole, to a minimum of 0.0126 m/s$^2$ at the equatorial locations of its major axis, due to both Haumea's shape and the strength of Haumea's centrifugal acceleration. Furthermore, Haumea exhibits a non-monotonic decrease in $g$ with latitude, along with strong $g_{\theta}$ terms that result in  Haumea's surface gravity vectors pointing poleward relative to the surface normal at higher latitudes ($>$ 60$^{\circ}$).

  \item The simple to complex crater transition diameter on Haumea is expected to vary greatly as a function of latitude. Using the observed transitions on icy bodies as a function of gravity we infer a simple to complex transition diameter ($D_t$) of 36.2 km at Haumea's location of minimum surface gravity at the equator, compared to $D_t$ = 6.1 km at the poles. Thus craters are more likely to be simple at equatorial regions, and complex in polar regions. In distinguishing whether the physics of the impact process is dominated by Haumea's surface material strength or surface gravity, the transition from the strength to gravity regime occurs at $\sim$0.5 km at the pole, and 150 $km$ at the equator.

  \item Due to the spatial variations in Haumea's surface gravity, an impactor of the same size and impact velocity will form craters with different morphometries across Haumea's surface. For craters in the gravity regime (larger craters), craters near the equator will be of larger volume, larger diameters, and considerably deeper than craters at mid-latitudes, followed by craters at the pole. Craters in the equatorial region are also less likely to be complex.

  \item These same spatial variations in crater characteristics for the same impactor will also extend to the ejecta, in the case of craters in the strength regime (smaller craters). Crater ejecta are expected to be thinnest at the location of maximum gravity at the poles, with thicknesses up to 10$\times$ higher at other locations on the surface, as well as up to 63$\times$ thicker in the immediate vicinity of the location of the major axis at Haumea's equator.

  \item Haumea's escape velocity varies by 38\% strictly across Haumea's equator, due to its shape as well as large angular velocity. The highest escape velocity at the pole (0.97 km/s) is 62\% more than the minimum equatorial escape velocity (0.60 km/s).

  \item Despite Haumea's escape velocity being higher at the poles, the larger minimum ejecta velocity ($v_*$) calculated for Haumea's higher latitudes result in a higher mass fraction of ejecta escaping Haumea's gravitational well at polar vs equatorial latitudes for impactors of the same size.
\end{enumerate}

Our predictions for Haumea reinforce the notion that despite broad similarities in bulk densities, comparable radii, and occupying the same broadly defined region of our solar system, the large Kuiper Belt Objects are incredibly diverse and worthy of further individual investigation. Haumea does not possess the larger volatile inventory of Pluto or Eris, but the spatial varitions in its crater characteristics may be unseen elsewhere in the Kuiper Belt, as well as in the solar system.

\section*{Acknowledgements}

Support for GDM and LO was provided by a startup grant from Rutgers University.

\section*{Appendix: Full Expansion of Gravitational Terms}
\label{appndx_a}

We begin with the gravitational potential expressed as a series of spherical harmonics, the same relation as equation \ref{eqn1} in the manuscript:

\begin{align}
\begin{split}
  \label{eqn_Phi_sum}
  \Phi(r,\theta,\lambda) = - \frac{GM}{r} \biggl\{
                             1 & + \sum_{n=2}^{\infty} \sum_{m=0}^{n}
                             \left( \frac{R_o}{r} \right)^n P_n^m (\cos \theta) \\
                           \times [ & C_{nm} \cos m \lambda +  S_{nm} \sin m \lambda] \biggr\} \\
                           & - \frac{1}{2} \omega^2 r^2 \sin^2 \theta
\end{split}
\end{align}

evaluated explicitly up to n = 4, as we do for all calculations in the manuscript, this is:

\begin{align}
\begin{split}
  \label{eqn_Phi}
  \Phi = - \frac{GM}{r} - \frac{\alpha_r GMR_o^2}{r^3} - \frac{\beta_r GMR_o^4}{r^5}
              - \frac{1}{2} \omega^2 r^2 \sin^2 \theta
\end{split}
\end{align}

where

\begin{align}
\begin{split}
  \label{eqn_alpha_r}
  \alpha_r = \left( \frac{Ro}{r} \right)^2 \left[ \frac{C_{20}}{2} (3 \cos^2 \theta - 1) +
              3 C_{22} \sin^2 \theta \cos(2 \lambda) \right]
\end{split}
\end{align}

\begin{align}
  \begin{split}
    \label{eqn_beta_r}
    \beta_r = & \left(\frac{Ro}{r} \right)^4 \biggl[ \frac{C_{40}}{8} (35 \cos^4 \theta - 30 \cos^2 \theta + 3) \\
              + & \frac{15 C_{42}}{2} (7 \cos^2 \theta \sin^2 \theta \cos(2 \lambda) - \sin^2 \theta \cos (2 \lambda)) \\
              & + C_{44} 105 \sin^4 \theta \cos (4 \lambda) \biggr]
  \end{split}
  \end{align}

The surface gravitational acceleration is then calculated as the negative gradient of the gravitational potential (same relation as equation \ref{eqn2} in the manuscript):

\begin{align}
\begin{split}
    \label{eqn_g}
    \vec{g} = & - \vec{\nabla} \Phi \\
            = & - \frac{\partial \Phi}{\partial r} \hat{r}
                - \frac{1}{r} \frac{\partial \Phi}{\partial \theta} \hat{\theta}
                - \frac{1}{r \sin \theta} \frac{\partial \Phi}{\partial \lambda} \hat{\lambda}
\end{split}
\end{align}

The components of the gravitational acceleration are now explicitly evaluated, beginning with the $\hat{r}$ component $\vec{g}_r$:

\begin{align}
\begin{split}
    \label{eqn_g_r}
    \vec{g}_r = & - \frac{\partial \Phi}{\partial r} \hat{r} \\
            = & \left( - \frac{GM}{r^2} - \frac{3 \alpha_r G M R_o^2}{r^4}
            - \frac{5 \beta_r G M R_o^4}{r^6} + r \omega^2 \sin^2 \theta \right) \hat{r} \\
            = & ( g_{r,1} + g_{r,2} + g_{r,3} + \omega_{r} ) \hat{r}
\end{split}
\end{align}

Followed by the $\hat{\theta}$ component $\vec{g}_{\theta}$:

\begin{align}
\begin{split}
    \label{eqn_g_theta}
    \vec{g}_\theta = & - \frac{1}{r} \frac{\partial \Phi}{\partial \theta} \hat{\theta} \\
            = & \left( \frac{\alpha_\theta G M R_o^2}{r^4} +
                \frac{\beta_\theta G M R_o^4}{r^6} +
                r \omega^2 \sin \theta \cos \theta \right) \hat{\theta} \\
            = & (g_{\theta,1} + g_{\theta,2} + \omega_{\theta} ) \hat{\theta}
\end{split}
\end{align}

where

\begin{align}
\begin{split}
    \label{eqn_alpha_theta}
    \alpha_\theta = -3 C_{20} \cos \theta \sin \theta
                  + 6 C_{22} \sin \theta \cos \theta \cos (2 \lambda)
\end{split}
\end{align}

\begin{align}
\begin{split}
    \label{eqn_beta_theta}
    \beta_\theta = & \frac{C_{40}}{8} (-140 \cos^3 \theta \sin \theta +
                      60 \cos \theta \sin \theta ) \\
                   + & \frac{15 C_{42}}{2} [7 \cos(2\lambda)
                      (-2 \cos \theta \sin^3 \theta + 2 \cos^3 \theta \sin \theta) \\
                   - & 2 \sin \theta \cos \theta \cos (2 \lambda) ]
                   + 420 C_{44} \sin^3 \theta \cos \theta \cos (4 \lambda)
\end{split}
\end{align}

And finally the $\hat{\lambda}$ component $\vec{g}_{\lambda}$:

\begin{align}
\begin{split}
    \label{eqn_g_lambda}
    \vec{g}_\lambda = - \frac{1}{r \sin \theta} \frac{\partial \Phi}{\partial \lambda}
                            \hat{\lambda}
                    = & \frac{\alpha_\lambda G M R_o^2}{r^4 \sin \theta} +
                            \frac{\beta_\lambda G M R_o^4}{r^6 \sin \theta} \\
                    = & (g_{\lambda,1} + g_{\lambda,2}) \hat{\lambda}
\end{split}
\end{align}

where

\begin{align}
\begin{split}
    \label{eqn_alpha_lambda}
    \alpha_\lambda = - 6 C_{22} \sin^2 \theta \sin(2\lambda)
\end{split}
\end{align}

\begin{align}
\begin{split}
    \label{eqn_beta_lambda}
    \beta_\lambda = &-15 C_{42} (7 \cos^2 \theta - 1) \sin^2 \theta \sin(2\lambda) \\
                    & - 420 C_{44} \sin^4 \theta \sin(4\lambda)
\end{split}
\end{align}

\bibliographystyle{aasjournal}
\bibliography{library_23}

\end{document}